\documentclass[pra,amssymb,amsmath]{revtex4}

\usepackage{color}
\newtheorem{thm}{Theorem}

\newtheorem{defn}{Definition}

\begin{document}

\title{Quantum  randomness and free will}  
\author{Chetan S.   Mandayam Nayakar}
\affiliation{Poornaprajna Institute  of Scientific Research, 
Sadashivnagar, Bangalore.}
\author{R.  Srikanth} 
\email{srik@poornaprajna.org}
\affiliation{Poornaprajna Institute  of Scientific Research, 
Sadashivnagar, Bangalore.}

\begin{abstract}
Both deterministic and  indeterministic physical laws are incompatible
with  control  by  genuine  (non-illusory)  free will.  We  propose  that  an
indeterministic  dynamics  can be  {\it  weakly}  compatible with  free
will (FW),
whereby the latter acts  by altering the probability distribution over
allowed  outcomes.  In  the  quantum  physical world,  such  a FW  can
collapse the wave function, introducing deviations from the Born rule.
In principle, this deviation would stand in conflict with both special
relativity  and  (a  variant  of)  the  Strong  Church-Turing  thesis,
implying  that the  brain  may  be an  arena  of exotic,  non-standard
physics.  However, in practice, these deviations would not be directly
or   easily   observable,   because   they   occur   in   sub-neuronal
superpositions in  the brain, where  they would be shrouded  in random
measurement  errors, noise and  statistical fluctuations.   Our result
elucidates the difference  between the FW of human  observers and that
of observed particles in the  Free Will Theorem.  This difference is a
basic reason  for why  FW (and, in  general, consciousness)  cannot be
recreated  by standard  artificial intelligence  (AI)  technology.  We
propose  various  neurobiological  experiments  to test  our  proposed
theory.  We  speculate that  for observers to  be aware of  a physical
theory such as quantum mechanics,  FW is necessary and that the theory
must therefore not  be universal.  We suggest that  FW may be regarded
as   a  primitive   principle   in  Nature   for  explaining   quantum
indeterminism.

\end{abstract}

\maketitle

\section{Introduction\label{sec:intro}}

Free will  (FW), as we  normally understand it  in daily life,  is the
power to make one's own  choices.  Our outlook on the world implicitly
assumes that human behavior is governed  by FW: we choose, we plan and
we normally hold people responsible for what they say or do. Recently,
a  number  of  physicists  have  studied  FW in  the  context  of  the
foundations of  quantum mechanics (QM), in particular,  in relation to
non-classical  features like indeterminism,  entanglement, non-realism
and                                                       contextuality
\cite{ck,stapp,bas07,tum07,hoo07,nic,suControl,vedral,gisin}.

Determinism means that the past history of the universe 
completely determines the future of a
system, leaving no room for  FW to manoeuvre it.  Hence, 
unless FW is assumed to be illusory, deterministic
laws  of  physics  are  logically  incompatible  with  the  action  of
free-willed agents.  Neither is  randomness good for FW.
Evolution is random if the  physical laws are indeterministic, so that
the past  history of  the universe does  not determine  completely the
future evolution  of a system.  Nevertheless,  this ``freedom" will
be governed  by some probability rule,  like the Born  rule of quantum
mechanics  (QM).   If  there  is  no rule  to  choose  from among  possible
alternatives, one would still expect a default, democratic rule of all
alternatives being equally probable or an appropriate noise pattern.  
The frequencies of
outcomes would be  constrained by the law of  large numbers applied to
the  relevant  rule.   This  would  exclude the  possibility  of  full
control, and hence of FW as we normally understand it.  Therefore, the
``free" (or ``random") aspect of FW stands in conflict to the ``will''
(or ``control'') aspect, making the word ``free will" an oxymoron.
We call this the weak FW paradox (WFWP).

Informally, WFWP says  that if we accept FW  axiomatically, then FW can
coexist  peacefully  neither  with  determinism nor  randomness  in
physical  reality. A  strong version  of  the FW  paradox (SFWP)  asks
whether,  in this  light,  such a  thing as  FW  exists at  all or  is
definable.  
Suppose we wish to argue that a quantity $x$ (say, the position of
one's hand) is free-willed. To make the case that $x$ is
not random, but determined entirely by one's voluntary decision, we write
$x = x({\bf w})$, where ${\bf w}$ is the `intention variable'.
To ensure that $x$ does not thereby become deterministic, we require
that ${\bf w}$ is free and not determined by the past history of
the universe. But this would ${\bf w}$, and by consequence $x$, as 
random. So we have to assume that ${\bf w}$ itself, like $x$, is
determined by other `lower level' variables. An unending regression of
this sort is obvious. From this standpoint,  
determinism and  pure
randomness seem to  be the only fundamental primitives  to construct a
dynamics; FW as yet another  primitive does not seem to exist,
or at  rate, seems to be  undefinable in the  conventional language of
physics. SFWP alerts us to the fact that if FW exists, 
then it must be sought in
the middle-ground between randomness and determinism, and this, in a
way to be made rigorous below, will be our approach.
  
The remaining article is arranged as follows.  In Section \ref{sec:R},
we  present  a possible  resolution  of WFWP,  based  on  a clue  from
Tarski's semantic theory  of truth, and a partial  solution to SFWP, a
more  detailed   treatment  of  which  will  be   taken  up  elsewhere
\cite{yyz}.  The  implication of  our work for  the Free  Will Theorem
\cite{ck}  is discussed  in  Section \ref{sec:FWT}.   The question  of
compatibility with  relativity and efficient  computability is studied
in Section  \ref{sec:STR}.  In  Section \ref{sec:ex}, we  consider the
neuroscientific implications  and  
experimental   tests   of  our   proposal.
Philosophical  and  metamathematical  ramifications  are taken  up  in
Section \ref{sec:godel},  before concluding in the  final section.  We
will  refer to  WFWP and  SFWP  together as  ``FWP''.  Throughout  the
article,  we  will  find  it  convenient to  adopt  a  Copenhagenesque
interpretation  of QM  and talk  of ``wavefunction  collapse", ``state
vector   reduction",  "the  quantum   measurement  problem"   and  the
quantum-classical   divide,  mainly   because   this  terminology   is
convenient  for  our  purpose   and  is  used  informally  by  working
physicists.  Any  imprecision  in   the  usage  does  not  affect  our
arguments.  We indicate our preferred position in this regard later.

\section{The free will paradox and consequences\label{sec:R}}

FWP  does not  arise if it  assumed that  FW is illusory  and that
choices are made deterministically in the subconscious and uploaded to
the  conscious mind  that is  unaware that  it lacks  true  freedom to
choose \cite{hoo07,nic}.   For our purposes, by ``FW",  we will always
mean  genuine--  and not  illusory--  FW,  unless explicity  otherwise
qualified.
Another  resolution  to  WFWP   \cite{suControl}  could  be  that,  though
conforming to  the required probability  distribution, FW has  room to
alter  the  {\it  ordering}  of  outcomes.  This  would  require  that
free-willed  outcomes  (say  a  stream  of  bits)  would  have  to  be
compensated by  uncontrolled outcomes at other  times to restore
eventual conformance with the probability distribution.

Our proposed  resolution of FWP is  based on an insight  due to Tarski
for resolving the liar paradox \cite{tarski}, closely related
to  G\"odel's incompleteness  theorem \cite{god}.   The  liar paradox,
which  self-referentially  states {\it  this  statement  is false}  is
inconsistent because it is true if  and only if it is false.  To avert
the paradox,  it is necessary  to distinguish between the  langauge of
discussion  (called the  {\it object  langauge} $L^0$)  from  the {\it
  meta-langauge} ($L^{1}$),  in which to talk {\it  about} $L^0$.
The semantic truth of statements in $L^0$ cannot consisently be asserted
in  $L^0$ itself,  but  in $L^{1}$.   It  is the  failure  to make  this
distinction between $L^0$ and $L^{1}$ that leads to the liar's paradox.

Analogously,  in the  present situation,  we posit  a basic  or zeroth
level of physical reality,  called {\it objective reality} and denoted
${\bf R}^0$,  the arena where  the standard laws of  quantum mechanics
govern  physical or  {\it objective} quantities  and variables like  the
position  and momentum  of  particles. This  is  distinguished from  a
meta-level or first level  of reality, called {\it subjective reality}
and denoted ${\bf  R}^{1}$, the arena where some  other laws of Nature
govern  certain {\it  subjective} quantities  and variables,  like FW,
emotions and thoughts of sentient beings like human agents.

Any procedure outputing an  {\it objective} result (one describable in
the   language   of   standard   quantum   physics)   is   called   an
objective-valued (OV) procedure.  An experimenter selecting a detector
setting,  or a computer  printing out  a read-out  are examples  of OV
procedures.   We can  consistently  characterize FW  at the  objective
level as follows.  An OV action is said to be free-willed if (a) it is
not completely determined  by the past {\it objective}  history of the
universe;  and (b)  it  depends explicitly  on  some {\it  subjective}
variables.

Condition (a) implies that the outcome  of the action is random at the
objective level,  or ${\bf R}^0$-random.  Condition (b)  leaves room for
the subjective influence/control  of the action.  FW can  thus be said
to simultaneously  and consistently have  both freedom and  control by
recognizing that  the freedom  is with respect  to ${\bf  R}^0$, whereas
control is with respect to  ${\bf R}^{1}$. The distinction between the
subjective and objective realities is, in this way, the key to solving
WFWP, and  highlights the fact  that it is  not possible to  define FW
acting  on ${\bf  R}^0$ {\it  within} standard QM,
the  language/dynamics  of ${\bf   R}^0$.

For our present  purpose, a physical particle is taken  to be a purely
objective entity, living  in ${\bf R}^0$; by contrast,  a human being,
such as  an observer  measuring a photon's  polarization, is  a
subjective-objective entity.   In particular, she/he  is characterized
by subjective  degrees of  freedom (specifying thoughts, feelings
and FW) that dynamically couple  in some so-far unknown way to certain
objective  degrees of  freedom in  the brain.  In consonance  with the
terminology of Cartesian dualism,  we call the subjective component as
her/his  {\it mind}, and  the objective  component her/his  {\it body}
\cite{mbp}.  What we  have called  `objective' and  `subjective', are,
traditionally in classical  metaphysics, referred to as `physical'
and `non-physical', respectively.

In the sense specified, a particle has no mind. The `FW' of a particle
can only satisfy condition (a),  but not (b).  It has freedom but
no control.  The particle's FW is simply pure randomness. FWP thus
does  not  arise   in  the  case  of  particles.    To  normalize  the
terminology, we will call this particulate FW as {\it zeroth-order} FW
(zeroth-order freedom and no control),  in contrast to the FW of human
agents,  which  satisfies  both   conditions  (a)  and  (b)  and  will
henceforth be  called {\it  first-order} FW (first-order  control over
zeroth-order freedom).  An entity with first-order FW necessarily also
has zeroth-order  FW, as  it can choose  to run  or not run  a quantum
random process, (e.g., measuring $\sigma_x$  on a qubit prepared in an
eigenstate  of $\sigma_z$) as  a subroutine.   (In fact,  human agents
possess second-order FW, a point we return to elsewhere \cite{yyz}.)
To avoid repetetiveness, in the remaining article, we sometimes drop
the qualifications `first-order' or `zeroth-order' when the context
makes it clear which of these two is meant.

It remains to understand  the mechanism of
how first-order FW influences objective
degrees of  freedom in physical reality. This  will automatically help
solve the  mind-body problem  of dualism, which  is the absence  of an
empirically identifiable {\it mind-body interface}, or a meeting point
between  the  mind  and  body.   The  key  observation  here  is  that
probabilistic  laws governing ${\bf  R}^0$ will  in general  be compatible
with FW only if the control  feature (b) is absent.  This implies that
first-order  FW,  which incorporates  (b),  will  couple to  objective
degrees   of   freedom   {\it   by   causing   deviations   from   the
  otherwise-expected objective probability distributions}.

An illustration:  suppose $X$  is an OV  random variable
with  some probability  distribution  $P$, and  is  influenced by  the
free-willed  intervention of some  agent.  Given  any sequence  of $n$
tosses    of    $X$,     represented    by    $\underline{X}    \equiv
X_1X_2,\cdots,X_n$,   such   that   the   probability   of   occurance
$P(\underline{X}) >  0$, then  FW can force  the occurance  of outcome
$\underline{X}$ without {\it logical}  contradiction.  That is, it can
realize   any   possible    evolutionary   branch   allowed   by   the
indeterministic evolution rules of  the system, but cannot realize any
not allowed  by the rules.  By  controling the evolution  to be within
the space  of allowed branches, FW  does not stand  in direct conflict
with the  rule specified by $P$, in  a sense, even though  in the long
run, such FW-induced interventions  will become manifest as departures
from  $P$.   We  call  this  relation  between  FW  and  $P$,  whereby
free-willed  interventions  contravene   $P$  statistically,  but  not
logically, as the {\it weak compatibility} of FW and $P$.

In  specific,  if  $X$  is  the random  variable  corresponding  to  a
FW-influenced quantum  measurement ${\cal M}$ on system  $S$, then $X$
may deviate from the Born  probability rule (that this rule represents
genuine freedom of particles, under certain reasonable assumptions, is
shown  in Section  \ref{sec:FWT}). For  example,  if $S$  is a  qutrit
(three-level  quantum  system)   in  the  state  $|\psi\rangle  \equiv
\cos\theta|0\rangle +  \sin\theta|1\rangle + 0|2\rangle$,  then FW can
control and direct the collapse  of $|\psi\rangle$ under ${\cal M}$ to
yield $|0\rangle$ or $|1\rangle$ with  certainty, or at any rate, with
probabilities   other  than   the  standard   Born   probabilities  of
$\cos^2\theta$  and  $\sin^2\theta$,  respectively.  However,  outcome
$|2\rangle$  is forbidden by  virtue of  weak compatibility.   FW will
therefore manifest as {\it  statistical deviations from the Born rule}
over allowed  outcomes.  By contrast, zeroth-order FW,  which is plain
quantum randomness, will conform to the Born rule in the long run.

From an  objective perspective, the wavefunction  is {\it intangible}.
It  cannot be manipulated  or controlled  in ways  that depend  on its
value.   For example,  an amplitude  cannot be  abruptly set  to zero.
Mathematically,  this makes  QM  a linear  theory.   By contrast,  the
FW-directed collapse  is a nonlinear  phenomenon. This makes
the  wavefunction a subjective  {\it tangible},  as manifested  by the
control FW has over it.  To  wit, one may think of the wavefunction as
a ghost that is physically insubstantial, but quite substantial in the
subtler  sub-physical  world.  The  wavefunction can  then  be  fancifully
described as  an informational ghost  that bridges the  subjective and
objective worlds across the mind-body interface.

Presumably, FW  is exercised as an observer's  subjective control over
the collapse of certain sub-neuronal superposition states in the brain
under special conditions available there.  It is assumed
that these  conditions also determine the specific  basis of collapse.
Subjective  degrees of freedom  (characterizing the  agent's volition,
thoughts,  etc.)  couple  dynamically in  some so-far  unknown  way to
objective   degrees   of   freedom   (pertaining  to   motor   neurons
deterministically associated with  various voluntary bodily functions)
through this mind-body interface.  We note that though the exercise of
FW  is conscious,  the mechanism  by  which it  controls state  vector
reduction, and  its subsequent deterministic
amplification  to macroscopic classical
actions, will itself be at a sub-conscious level.

We  will  find  it  convenient,  following the  practice  in  Vedantic
philosophy,  to  identify  the  mind  as  the  subjective
counterpart of  the brain areas corresponding to  emotions, memory and
FW, and  identify the {\it  intellect} as the subjective  component of
the ratiocinative  and determinative faculty  in the brain \cite{veda}.  
The  model of
free-willed  action  that we  propose  is  the following  three-staged
process:

\begin{description}

\item[Stage 1-- Attention.] Faced  with a situation that requires
  a choice to  be made from among various  alternatives $j$, the brain
momentarily creates   a sub-neuronal 
quantum   superposition   $|\psi_\alpha\rangle  =   \sum_j
  \alpha_j|j\rangle$,   which  is  presented   to  the   mind.   
(Based on the idea presented in 
Ref. \cite{penham}, brain microtubules may be the seat of
such superpositions.) The   coefficients $\alpha_j$ of 
$|\psi_alpha\rangle$ may reflect certain priorities set
  by the physical brain.

\item[Stage 2-- Selection.] The  mind transfers the quantum
information $|\psi_\alpha\rangle$ to the
  intellect  (Why  this intermediate  step  is  required is  explained
  elsewhere  \cite{yyz}.)  The  intellect evaluates  the value  of the
  alternatives $j$ according  to  some  norm (emotional,  social,  etc.),
  choosing the  optimal one, $J$, from among  them. If the  norm does not  enable the
  choice of one  of the alternatives, a  random choice could  be made (by
  invoking a  neural zeroth-order FW subroutine).   That the selection
  may be influenced,  but is not determined, by  the amplitudes of the
  superposition,  is  at  the  origin  of  departures  from  the  Born
  rule. The intellect conveys the choice $J$ to the mind.

\item[Stage 3-- Collapse.]  The mind  exercises FW to direct the 
quantum state
  $|\psi_\alpha\rangle$ to  collapse the quantum superposition to
  the state $|J\rangle$ corresponding to the chosen alternative.
\end{description}

We   refer  to   this   model   as  the   ``ASC"   (the  initials   of
``Attention-Selection-Collapse") model. Because the selection in stage 2
is not according to the Born rule, stage 3 will in general cause a 
violation of energy conservation: a measurement described by
projective operators $M_j$
non-selectively transforms the density 
operator $\rho$ according to
$\rho \longrightarrow \rho^\prime \equiv \sum_j M_j\rho M_j^\dag$,
assuming the Born or trace rule for probability.
This transformation preserves energy, i.e.,
${\rm Tr}({\cal H}) = {\rm Tr}({\cal H}\rho^\prime)$
if the measurement operator ${\cal M} \equiv \sum_j m_j M^\dag_jM_j$ 
commutes with the system Hamiltonian ${\cal H}$. This is the case for
quantum non-demolition (QND) measurements \cite{qnd} but in general not
true, when the interaction is dissipative and
energy is exchanged with the environment \cite{qnd}. 

Even if we assume, conservatively, that the basis in which
FW-induced collapse happens in the energy basis
and thus commutes with ${\cal H}$, the 
Born deviation implies that in general, energy will not be conserved
on average. This is not a cause for worry because the violation of
energy conservation would  occur in a small sub-cellular region
of the motor cortex of the brain, where it will be hardly
discernible from measurment errors and 
the surrounding nervous noise. Moreover,
the remaining features of the brain's physiology can be
described in terms of deterministic observable
causal chains obeying the conventional physical conservation laws
(e.g., the metabolism involved in the arousal
potentials triggering bodily movements) \cite{suControl}. 
This fact also makes observing FW-induced effects directly
difficult (cf. Section \ref{sec:ex}).

The  immediate main gain from
the ASC model is  that it  allows  us to  formally distinguish  inanimate
entities like  particles from  animate entities like  human observers,
from the viewpoint of their degree of freedom and conscious initiative
or control.
To illustrate this we will consider two classes of entities, which are
externally  or objectively similar,  but are  distinguishable formally
according the above model.
\begin{defn}
A Constantly Good Person (CGP)  is an entity that, when presented with
a good and bad alternative of action, evaluates their moral value, and
exercises his FW to select the good alternative.
\label{def:cgp}
\end{defn}

We define  a binary moral-valued  function ${\cal M}({\bf x})$  by $f:
\Xi  \mapsto$  \{``good",``bad"\},  where   $\Xi$  is  the  space  (of
description in some language) of  actions in real-life contexts. It is
straightforward  to  generalize  this  to a  real-valued  moral-valued
function where, in some fashion, the moral value is quantified.

\begin{defn}
A Noble  Robot (NR) is a (quantum) algorithm that, when presented  with a good
and  bad alternative  of  action, computes  their  moral value  ${\cal
  M}({\bf x})$, and selects the good alternative.
\label{def:nr}
\end{defn}

In objective terms and the langauge of standard QM, a CGP is difficult
to  distinguish from an  NR.  As  the mind  is an  unobservable mental
agency affecting  the random dynamics of objective  variables, and the
quantum wave  function is intangible,  the ontological status  of both
these  items remains  unclear in  standard QM.   At best  we  may find
certain  neural correlates in  the CGP's  brain suggestive  of genuine
moral judgment, but we can never  conclusively argue that he is not an
NR  of some  sort.   However, it  is  clear that  the richer  langauge
required in the  ASC model entails that the  two systems have distinct
descriptions  at a  subjective level.  This forms  the content  of the
following theorem.
\begin{thm}
A CGP and NR are first-order distinguishable.
\label{thm:cgpvsnr}
\end{thm}
{\bf Proof sketch.}  When a  NR's logical processor computes the moral
value  of  the presented  two  alternatives,  only the  ``good"-valued
alternative triggers  a switch that  controls the NR's  motor cirtuit.
By contrast, presented with the  alternatives, a CGP's brain creates a
quantum superposition of the  alternatives (stage 1).  CGP's intellect
determines the good alternative, and the mind directs the wavefunction
collapse to  the state corresponding  to this alternative.  The  NR is
algorithmically   and   dynamically   bound   to   choose   the   good
alternative. The idea  that CGP is not bound by  the objective laws of
physics and freely chooses the  good alternative is represented by the
superposition  state, which  implies  that the  CGP  {\it could}  have
chosen otherwise but {\it would} not.  \hfill $\blacksquare$

Theorem  \ref{thm:cgpvsnr}   can  be  extended  to   indicate  why  AI
technology, even  with access  to standard quantum  randomness, cannot
recreate FW. Although an AI algorithm for an NR can, to a good extent,
simulate a CGP, it is qualitatively different. More generally, one can
define  an {\it  inconsistently  noble robot},  where a  probabilistic
selection  procedure  involves the  creation  of  a superposition  and
measurement  in  the  computational   basis.  By  constrast,  an  {\it
  inconstantly  good  person}  would  require  quantum  dynamics  with
deliberate, subjective intervention.  These considerations entail that
a  human agent,  and by  extension any  sentient agent,  could  not be
considered merely as a sufficiently complex robot, but a qualitatively
distinct  class of  entities.   Here Penrose's  interesting thesis  is
worth noting, according to which conscious processes are fundamentally
non-algorithmic \cite{pen94}, an idea that is implicit in ancient East
Asian philosophies.

\section{Implications for the Free Will theorem \label{sec:FWT}}

Zeilinger notes  that there are two  ``freedoms": that on  the part of
Nature, and that on the  part of the experimental observer \cite{zei}.
Echoing  this idea more  rigorously, the  FW theorem  (FWT) \cite{ck},
which is essentially a non-locality proof for any theory that purports
to reproduce  quantum correlations, states that if  observers have FW,
then so too do observed particles.  A brief account of FWT, based on a
test of quantum nonlocality given  in Ref.  \cite{kar}, is as follows.
Consider 9  orthogonal bases  in a Hilbert  space of dimension  4, and
denote  them by $S_1,  S_2, \cdots,  S_9$. Each  $S_j$ is  a set  of 4
orthogonal vectors, denoting directions of detector settings, which we
write   in   un-normalized,   `row'  representation   as   $(1,0,0,0),
(1,{-1},0,0) \equiv \frac{1}{\sqrt{2}}((1,0,0,0) - (0,1,0,0))$, etc.

\begin{center}
\begin{table}
\begin{tabular}{|l|l|l|l|l|}
\hline
$S_1$  & $(0,0,0,1)$ & $(0,0,1,0)$ & $(1,1,0,0,)$ & $(1,{-1},0,0)$ \\
$S_2$  & $(0,0,0,1)$ & $(0,1,0,0)$ & $(1,0,1,0)$ & $(1,0,{-1},0)$ \\
$S_3$  & $(1,{-1},1,{-1},)$ & $(1{-1}{-1}1)$ & $(1,1,0,0)$ & $(0,0,1,1)$ \\
$S_4$  & $(1,{-1},1,{-1})$ & $(1,1,1,1)$ & $(1,0,{-1},0)$ & $(0,1,0,{-1})$ \\
$S_5$  & $(0,0,1,0)$ & $(0,1,0,0)$ & $(1,0,0,1)$ & $(1,0,0,{-1})$ \\
$S_6$  & $(1,{-1},{-1},1)$ & $(1,1,1,1)$ & $(1,0,0,{-1})$ & $(0,1,{-1},0)$ \\
$S_7$  & $(1,1,{-1},1)$ & $(1,1,1,{-1})$ & $(1,{-1},0,0)$ & $(0,0,1,1)$ \\
$S_8$  & $(1,1,{-1},1)$ & $({-1},1,1,1)$ & $(1,0,1,0)$ & $(0,1,0{-1})$ \\
$S_9$  & $(1,1,1,{-1})$ & $({-1},1,1,1)$ & $(1,0,0,1)$ & $(0,1{-1},0)$ \\
\hline
\end{tabular}
\caption{List of 9 observables Alice can measure, out of 18 possible
directions.}
\label{tab:jhon}
\end{table}
\end{center}

Measuring $S_j$ on  a state that is not  its eigenstate, will collapse
it to an outcome randomly  according to the Born rule. This randomness
by  itself  does not  at  first imply  the  freedom  of the  particle.
Measurement only  {\it finds} the outcome random.  Unknown causes from
the past may  well have made it determinate  prior to measurement, the
randomness being  an artefact  of our not  knowing those  causes.  The
Kochen-Specker theorem  \cite{kc} rules out that  any such determinate
assignment  of  outcomes  exists  independent  of  context  (here  the
$S_j$'s).

A  proof of  the  Kochen-Specker  theorem is  as  follows.  Suppose  a
determinate  assignment  exists for  each  vector  appearing in  Table
\ref{tab:jhon}. Thus  in each  row of Table  \ref{tab:jhon}, precisely
one  of the  four  vectors is  to be  assigned  the value  1, and  the
remaining three the  value 0 (corresponding to the  SPIN assumption of
FWT).  The same value (0 or 1) is to be assigned to all instances of a
vector: eg.,  (1000) is  assigned the same  value in $S_1$  and $S_2$.
But a careful  look at the Table shows that  no such assignment scheme
exists.  For the  sum of assignments across each row  is 1, yielding 9
as  the sum  of  values for  each  row. But  then  each vector  occurs
precisely  twice  over all  rows  taken  together.   Thus the  sum  of
assignments  of   all  vectors  should   be  an  even   number.   This
contradiction implies  that no determinate assignments can  be made to
the vectors, except if they are context-dependent. This means that two
instances of the same vector,  e.g., $(1,0,0,0)$ occuring in $S_1$ and
$S_2$, need not have the same value.  This impossibility to pre-assign
values means either that particles have freedom, or their values could
be predetermined in a context-dependent way.

Therefore, to  show that particles have  freedom, we need  to create a
situation where  information about  context will not  available during
measurement. To  this end consider Alice and  Bob sharing entanglement
of the form:
\begin{equation}
|\Psi_J\rangle = \frac{1}{2}
\sum_k |S^{(k)}_J\rangle_A|S^{(k)}_J\rangle_B,\hspace{1.0cm}
(J=1,\cdots,2).
\end{equation}
It can be shown that this form is preserved in any of the above bases,
i.e., $|\Psi_1\rangle = |\Psi_2\rangle = \cdots = |\Psi_9\rangle$.
This ensures that if Alice and Bob measure in the same basis $S_{j}$,
they will find the same outcome $k$ (corresponding to 
the TWIN assumption of FWT).

Alice randomly measures one of the 9 bases, $S_j$, which will serve as
context.  Denote  her  outcome  as $|k^\prime\rangle$,  to  which  she
assigns  value  1,  and  0  to  the  other  elements  of  $S_j$.   Bob
independently and randomly chooses one of the 18 vectors $k$ appearing
in   Table   \ref{tab:jhon}  and   measures   the  binary   observable
$|k\rangle\langle k| -  (\mathbb{I}-|k\rangle\langle k|)$.  He assigns
$|k\rangle$ the  value 1 or  0 depending on  whether he detects  it or
not. QM  guarantees that  if Bob's  vector $k$ happens  to lie  in the
context  $S_j$ they  will  assign it  the  same value  0  (in case  of
non-detection of $|k\rangle$) or 1  (in case of detection).  If we try
to explain  the above  in terms of  prior assignment of  values, Alice
should be able  to communicate her context information  $j$ to Bob. By
ensuring that Alice's and  Bob's meaurements are sufficiently close in
time, we can  rule out such context communication  at any finite speed
(FWT's  FIN  assumption,  later  weakened to  MIN).   If  relativistic
invariance  is assumed  (though  this is  not  necessary), then  their
measurement events are required to be spacelike separated.

We have thus far implicitly assumed that the observers have freely and
randomly selected their detector  settings.  It is still possible that
there is no  freedom of the particles, provided  that the experimental
settings  and the  outcomes are  pre-determined in  such a  way  as to
suggest stronger-than-classical  correlation, and  that the FW  of the
experimenters  is  an illusion.   (An  observation  that weakens  that
possibility   is   the  question   of   why   such  a   conspiratorial
hyper-determinism   should   go   only    so   far   as   to   suggest
stronger-than-classical  correlations  but  stop short  of  suggesting
superluminal  communication.)   Therefore,   if  we  assume  that  the
experimenters  have  genuine  FW,  then  all  avenues  of  attributing
determinate values to  Alice's and Bob's outcomes (no  matter what the
unknown  causes) are  ruled  out, and  they  must be  considered as  a
genuine acts of creation or FW on the part of the particles/Nature.

The  above  proof  of  FWT   is  not  suggestive  of  any  control  by
experimenters  on  the  randomness  they  generate. Their  FW  is  not
required  to be  any different  from the  zeroth-order FW  of observed
particles.  However, by virtue of Theorem \ref{thm:cgpvsnr}, we expect
the experimenters' FW to be of first-order.  We can envisage that (for
example)  the  experimenters  are  graduate  students  whose  work  is
affected by adverse comments on  their earlier run of trials, or plans
for a  exciting week-end mountaineering  trek. This could  manifest as
fluctuations  in  the  random   settings  they  chose,  which  may  be
detectable by  certain standard randomness tests.   This suggests some
kind of control  and hence that the FW  exercised by the experimenters
is  of first-order: the  `particle' in  the observer's  brain possibly
creates equal amplitude superpositions  for the alternatives.  But the
emotional vulnerability  of the mind  affects the choice of  the state
selected for collapse.

The  control  aspect of  the  FW of  human  agents  could manifest  as
large-scale  deviation from  the expectation  value of  some quantity.
For example, if the expected  behavior based on applying the Born rule
in stage 2 of the ASC model dictates following the local gradient in a
brain energy-potential landscape (e.g., to eat food placed in front of
a person), the person the brain  belongs to may exercise FW to deviate
from this  behavior (e.g., to fast  in the interest of  good health in
the long  run or in order to  feed an indigent person).   Thus, if the
control  aspect  of  FW  were  `switched off',  degrading  the  FW  to
zeroth-order, then a purely random, zombie-like behavior would result,
where agents' behavior will  lack any self-control, long-term planning
as  well as friendly/altruistic  behavior.  FWT  as it  stands neither
requires nor implies this difference  between the FWs of observers and
particles. Therefore, it does not preclude that human FW can be 
recreated with AI technology having access to standard quantum randomness.
Our distinction between the two orders of FW preclude this possibility.
Perhaps FWT  may be recast as a moral dilemma in
order to highlight the difference. We
will revisit this issue in a subsequent work \cite{yyz}.

\section{Implications for Special Relativity
and computational complexity\label{sec:STR}}

The laws of physics can be characterized by no-go conditions, which specify
operations that are physically impossible. Two such basic conditions
that have been recognized in quantum information science
are the no-signaling condition and the so-called ``the world is
not hard enough" (WNHE) assumption, 
an empirical rule-of-thumb based on a computer
scientific survey of the computational power of physical systems
\cite{sriwnhe,sri10}.
We expect them to be true because of, respectively, special relavitistic
restriction and our notion of what are reasonable models of computation.
As shown in the following two subsections, both
these no-go conditions are violated by non-Bornian probabilities,
essentially because the FW-induced effects make QM nonlinear.
However, as with energy non-conservation, they would happen
at the sub-cellular level within the brain. As a result,
they would be difficult to observe directly, and 
are certainly not contradicted by known empirical verifications 
of these conditions. In principle, however, these observations imply that
the brain is a seat of exotic and highly non-standard physics,
presenting a potential for uncovering new physics.

\subsection{Special relativity considerations}

In  a  multipartite  quantum  system, any  quantum  operation  applied
locally to  one part does not  affect the reduced  density operator of
the  remaining  part.   This  fundamental  no-go  result,  called  the
``no-signalling theorem'', implies  that quantum entanglement does not
enable   nonlocal    (``superluminal'')   signaling   under   standard
operations,  and is  thus  consistent with  special relativity.   This
peaceful coexistence of quantum nonlocality and relativity is arguably
tense \cite{gisin} because of  the violation of Bell-type inequalities
\cite{bell},  but  it  does  not overtly  imply  spacelike  influence.
Usable information always requires  the transport of a material object
and hence  cannot be communicated outside the  lightcone, in agreement
with relativity theory.

The ambiguity  between whether this spooky spacelike  signaling is for
real  or not, would  be broken,  and superluminal  communication would
become  a   reality,  if  measurement   outcomes  can  be   forced  on
superpositions,  in  violation of  the  Born  rule \cite{sri10}.   For
example,  if   agents  Alice  and   Bob  share  the   entangled  state
$\frac{1}{\sqrt{2}}(|00\rangle  + |11\rangle)$, and  Alice is  able to
harness FW  to deterministically collapse her qubit  to $|0\rangle$ or
$|1\rangle$, then she can control  whether to leave Bob's qubit in the
state $|0\rangle$ or $|1\rangle$,  respectively, and thus signal Bob 1
bit superluminally.   This fact remains  even if she has  only partial
control over  the collapse, so long  as there is a  deviation from the
Born  rule  (see  Ref.   \cite{sri10} and  references  therein).   The
signaling speed  attainable in this  manner is arguably the  `speed of
quantum information', namely  the lower bound on the  speed at which a
(superluminal) signal should travel  in order to explain the violation
of  Bell-type inequalities  \cite{bell}.  In  the  currently available
experimental estimate, this speed is at least of the order of $10^4 c$
\cite{salart}.

Therefore, if  we accept the  reality of FW, and  FW-induced collapse,
relativity must be considered  as an approximate description pertinent
only  to  {\it  objective   reality},  and  the  apparent  absence  of
superluminal signals is attributed  to the lack of macro-scale quantum
superpositions  with which subjective  variables can  interface.  From
the  perspective of  subjective  (`higher') reality,  space, time  and
causality  are  taken  to  be  absolute.  Subjective  quantities  like
first-order  FW  and states  of  mind  (emotions  and intentions)  are
frame-independent.

Consider  a Bell-type  experiment of  a pair  of  entangled particles,
where  two observers Alice  and Bob  freely and  independently measure
their respective  particle at spacelike-separated  events. Further let
Alice and Bob  be in relative motion such  that each considers her/his
own  measurement  as  the  first  \cite{bb}.  An  explanation  of  the
violation of Bell-type inequalities  in terms of a superluminal signal
traveling  from one  observer's measurement  event to  the  other's is
untenable  in  the relativistic  causal  description,  because of  the
incomparable time-ordering of the events.  In our formalism, since the
subjective, absolute perspective is taken to be the right picture, and
the  objective,  relativistic  perspective  the  approximate  picture,
invoking superluminal influence  to explain quantum correlations poses
no logical contradiction.  One of Alice  and Bob will be deemed as the
first to measure and the originator of the superluminal signal, though
this  fact makes  no difference  to an  observer who  interprets their
quantum correlations in the relativistic framework.

One might consider an experimental test  of our model of looking for a
inertial reference  frame, potentially  the absolute frame,  such that
simultaneous measurement  in that frame  leads to a break-down  in the
quantum  correlations of the  TWIN kind.   However, such  a break-down
need not  happen, if instantaneous communication  over finite distance
were  possible.   Such infinitely  fast  communication  would be  less
surprising if  we regard space,  not as a  physical barrier, but  as a
kind of information, and thus ultimately not unlike an internal degree
of freedom \cite{sri10}.

The very small, sub-neuronal  scale at which such FW-induced violation
of no-signaling occurs ensures that it is not invalidated by available
neurobiological data and experimental  tests in support of Relativity.
At the  same time, this  will present challenges for  its experimental
observation,  as  discussed in  Section  \ref{sec:ex}.  This  possible
FW-induced violation of  no-signaling is not in conflict  with the MIN
assumption  of FWT,  provided the  `speed of  quantum  information' is
finite. But  even otherwise, this mechanism  of superluminal signaling
presumes  FW  on the  part  of the  observer  and  particle, the  very
features FWT either assumes or purports to prove.

\subsection{Computational complexity considerations \label{sec:ccc}}

The central problem in theoretical
computer science  is the conjecture that the computational complexity
classes, {\bf  P} and  {\bf NP}, are  distinct in the  standard Turing
model  of computation.  ${\bf P}$  is the  class of  decision problems
efficiently (that is, in time that grows at most polynomially as a
function of problem input size) solvable by  a (deterministic) 
Turing machine (TM).  
${\bf  NP}$ is
the class  of decision problems  whose solution(s) can be  verified
efficiently by a deterministic TM.   
The word  ``complete" following a  class denotes a problem  $X$ within
the class, which is maximally hard in the sense that any other problem
in the  class can be  solved in polynomial time if there is an algorithm
that solves $X$ efficiently.
For  example, determining
whether  a Boolean  formula  is satisfied  is  {\bf NP}-complete.

{\bf P} is often taken to be the class of computational problems which
are  ``efficiently solvable"  (i.e., solvable  in polynomial  time) or
``tractable", although  there are potentially larger  classes that are
considered tractable such as {\bf BQP}, 
the class of  decision problems efficiently solvable by a
quantum  computer. {\bf  NP}-complete and  potentially
harder problems, which  are not known to be  efficiently solvable, are
considered intractable in the TM model.  If ${\bf P} \ne {\bf NP}$
and the universe is a polynomial-- rather than an exponential-- place,
physical  laws cannot  be harnessed  to efficiently  solve intractable
problems, and  {\bf NP}-complete problems  will be intractable  in the
physical world.

That classical physics supports  various implementations of the Turing
machine is  well known. More  generally, we expect  that computational
models supported by a physical  theory will be limited by that theory.
Empirical evidence suggests that the physical universe is indeed
a polynomial place \cite{scott}.
 We will informally refer to
the  proposition  that the  universe  is  a  polynomial place  in  the
computational  sense  as  well  as  the
communication sense by the expression  ``the world is not hard enough"
(WNHE)  \cite{sriwnhe}. This is closely related to what is sometimes
called the strong Church-Turing thesis, which asserts that any
"reasonable" model of computation can be efficiently simulated on 
a probabilistic Turing machine \cite{bz}.

Given a  black box binary function  $f(j)$ over $n$  bits, the problem
(called Boolean Satisfiability  or SAT) of determining if  there is an
input $j$  for which $f(j)=0$, is {\bf  NP}-complete. 
However, if FW  can collapse an
outcome   deterministically,  one   can  obtain   an   algorithm  that
efficiently  solves  {\bf  NP}-complete  problems.   One  prepares  in
polynomial      time     a     quantum      state $|\Psi^\prime\rangle$
as follows:
\begin{equation}
\frac{1}{2^{n/2}}\sum_j |j\rangle|0\rangle \stackrel{H}{\longrightarrow}
|\Psi\rangle \equiv \frac{1}{2^{n/2}}
\sum_j|j\rangle|f(j)\rangle,
\end{equation}
where the  second register  is a  qubit. One
then  uses   FW  to direct  the  second   register  to   collapse  to
$|1\rangle$. By virtue of weak compatibility, 
the  above decision problem  about $f(j)$ is  answered in
the  affirmative   if  and  only   if  the  second   register  returns
$|1\rangle$.

We conclude that WNHE is valid
in the objective world, when subjective or
sub-physical effects are excluded. Even otherwise, it is a very good
approximation, since the quantum coherence required to manifest
nonlinear violations of WNHE, are confined to sub-neuronal scales.
Further, within the objective scope, WNHE has good explanatory power
\cite{sri10}.
The very small scale of FW-induced contravention of WNHE also
present experimental challenges for its observation,
an issue taken up in Section \ref{sec:ex}.

\section{Neurobiological implications and 
experimental tests\label{sec:ex}}

Clearly,  FW-control of  collapse will  be  at least  as difficult  to
observe directly  as wave function  collapse itself.  Even  when brain
processes are  directly examined, FW-induced deviations  from the Born
rule  will hardly be  discernible from  measurement errors,  noise and
statistical fluctuations.  Any trace of FW will obviously be absent in
the macroscopic  `classical' world, where, for a  reason that famously
continues to be debated (but we  won't get into here), we don't get to
see quantum superpositions \cite{home}.  The challenge to neuroscience
technology will be  to tease out patterns attributable  to FW from the
frenzied buzz  of neuronal activity in  the brain. This  will no doubt
require  advanced  experimental  technology and  sophisticated  signal
processing.

We term the  idea that the effect of FW  remains veiled behind quantum
and classical  noise as the  {\it cognitive censorship}, to  wit, {\it
  the mind remains hidden in  matter}.  Because of this, the objective
universe appears  to be self-contained, with  no definitive indication
of the subjective  reality beyond, in the corpus  of known experiments
verifying  QM.  In  particular,  the  predicted  violation  of  energy
conservation, no-signaling and WNHE at the sub-neuronal level are, per
current  experimental  technology, are  barely  observable.  It is  an
interesting  question  whether   cognitive  censorship  will  make  FW
empirically unverifiable.   Here we take the  optimistic position that
FW  is  in   principle  not  be  beyond  the   scope  of  experiments.
In the remaining part of this Section, we point out how recent
neuroscientific experiments are in agreement with the ASC model,
and indicate possible ways of neurobiologically testing
our  proposed  model,  and   more  broadly,  the  idea  of  FW-induced
wavefunction collapse. \\

\paragraph{\bf Direct observation of neurons.}

Any spontaneous action of a  human or animal subject can presumably be
traced backwards along a  deterministic observable causal chain, e.g.,
from  nerve  endings in  the  muscles  controling  eye brow  movement,
through nerves  back to a specific  area in the motor  cortex.  In the
conventional view,  the brain is  a complex input-output  device, with
voluntary  actions arising  from external  stimuli.  
On  the other hand, the ASC  model implies that, though the brain
acts as an input-output processor, yet FW
makes the brain  a creative device that can  generate new information.
The output need not be entirely explained in terms of input.
Because the three stages of ASC require quantum coherence to
be present during their execution, ASC implies that the above causal
chain will originate from a single neuron-- rather than a multi-cellular
neural circuit-- in the motor cortex area of the brain. 
We term this the `opening neuron'.
Even if a neural circuit is recognized as closely associated
with the origin of a particular voluntary action, at its core there
will presumably be a single neuron that `wills' the action.

No doubt, identifying such single opening neurons will be difficult, given
the high density  of neuronal packing, and the  weakness of excitatory
synapses,  which would  make the  role  of individual  neurons in  the
mammalian  cortex hard  to  identify.  A  remarkable  study where  the
twitching of  a mouse's  whisker has been  traced to  single pyramidal
neurons in the cortex is reported in Ref. \cite{mouse}.
 
According to the  ASC model, the opening neuron is the  seat of FW and
new  physics. Experiments  may be  designed to  verify that  the cells
manifest  spontaneous, indeterministic  behavior such  as sub-neuronal
variations not  attributable to any  external stimuli, or  other, less
direct  signatures.  The  next step  will be  to use  neurobiology and
genetics  to  localize  and  understand  the  neuronal  organelle  and
concomitant brain  circuits responsible for  the spontaneous behavior.\\

\paragraph{\bf Brain scan studies through scalp electrodes,
fMRI, etc.}

Although studies based  on fMRI or scalp electrodes  cannot access the
quantum  regime, they  may be  able to  corroborate, modify  or reject
aspects  of the ASC  model.  Based  on a  study on  volunteers wearing
scalp electrodes, Libet and  collaborators \cite{libet} showed in 1983
that a `readiness potential' (RP)  was detected a few tenths of second
before  the subjects,  in their  own reckoning,  made the  decision to
perform an action (to flex a finger or wrist).  They interpreted their
result  as indicating  that the  motor  cortex was  preparing for  the
action,  that unconscious  neural processes determine  actions and
hence that FW was illusory.

In   a  recent  comment   on  the   experiment,  Miller   and  Trevena
\cite{trevena} asked subjects to wait  for an audio tone before making
a decision  to tap a  key or not.   If the activity detected  by Libet
really was the making of the decision prior to any conscious awareness
of doing  so, then that activity  ought to occur only  if the subjects
decided  to act.   But  no  such correlation  was  found.  Miller  and
Trevena  conclude that  the RP  may only  indicate that  the  brain is
paying attention and does not indicate that a decision has been made.

In the  light of the ASC  model, we can interpret  such experiments as
follows.  The  RP corresponds  to stage 1  (at time $t_1$),  where the
brain pays attention, creating a superposition of the form:
\begin{equation}
|\Phi\rangle \equiv
\alpha|{\rm tap~key}\rangle + \sqrt{1-\alpha^2}|{\rm don't~tap~key}\rangle,
\hspace{1.0cm} 0 \le |\alpha| \le 1.
\end{equation}
which is presented to the mind. 
At time $t_2$, the subject's intellect makes a decision to
select one of the two alternatives (stage 2). At a slightly later time, $t_3$,
the mind collapses the state vector $|\Phi\rangle$ to the chosen
alternative (stage 3). Our model thus explains the time gap between
the RP blip ($t_1$) and eventual decision ($t_2$) in these
experiments, and is compatible with the interpretation of
Trevena and Miller \cite{trevena} that the state at the RP stage is 
independent of the eventual decision made. \\

\paragraph{\bf Macroscopic considerations}

Although the  deviations from the  Born rule are  microscopic effects,
they  are  cascaded  into  macroscopic  actions  of  human  observers,
becoming amplified by nonlinear  chaotic processes in the brain.  This
is similar to the situation that  though the collapse of photon to one
of two paths in a beam splitter is not directly observable, its effect
amplified to classical levels is  visible as the detection by a simple
(avalanche) detector. Thus,  the ASC model entails that  the effect of
FW in sentient  beings will be an irreducible  residual 
behavior  that  cannot  be  accounted  for  by  external  stimuli.  
There will be variability in this residual behavior,
implying  that  the  brain  cannot  be  modeled  as  a
complex  input-output  device,  but  as a  source  of  information
creation. Unfortunately, demonstrating such variable, spontaneous behavior
experimentally
can be difficult, since animals' responding differently even to the
same external stimuli would normally be attributed to random errors
in the complex brains.

The ASC model implies, however, that because animals should in fact
possess first-order FW (however rudimentary), subjective variables
responsible for free-willed behavior will cause the above
variability to depart from a random noiselike pattern.

In  remarkable research  reported in  Ref. \cite{brembs},  fruit flies
(Drosophila  melanogaster) were tethered  in completely  uniform white
surroundings and  their turning behavior recorded. In  this setup, the
flies do  not receive any visual  cues from the  environment and since
they are fixed  in space, their turning attempts  have no effect. Thus
lacking  any  input,  their  behavior should  resemble  random  noise,
similar  to a  radio detuned from any station.   However,  the analysis
showed that the  temporal structure of fly behavior follows
a L\'evy-like probabilistic behavior pattern 
rather than resemble random noise, in conformance
with what the ASC model suggests.

Further tests would be to subject such flies to uniform external
stimuli, such as soft music, and monitor the variability. We expect
that such mild stimuli will not directly affect the brain region
responsible for the above motions, and thus not affect variability
pattern if its origin were random errors in the brain. On other
the other hand, the music will presumably affect the subjective state
of the flies. Therefore, if the pattern has its origin in 
subjective variables, we can predict a corresponding change in
the variability's pattern. \\

\paragraph{\bf Computational complexity approach}

Another approach, based on the results of Section \ref{sec:ccc},
is to study whether there are 
subconscious activities in the brain that correspond to problems
that are suspected to be intractable
from the (quantum) computation theoretic perspective, and yet
the brain demonstrably solves them efficiently. As the
ability to control wave-function collapse would bestow such 
super-Turing power (as noted in Section \ref{sec:STR}), the discovery of such
cognitive processes would support the ASC model.

For example, there are several intractable problems in pattern
recognition \cite{simon}. Humans, and perhaps even animals,
are known to recognize patterns, such as faces, in new and unfamiliar 
situations quickly and far more reliably 
than the best computer algorithms to date. If these real life
instances of subconscious
problem solving can be reduced to {\bf NP}-hard, and it can be
experimentally verified that it is done efficiently by human
subjects, this would serve as evidence of the ASC model.

\section{Whence free will? \label{sec:godel}}

Why does  FW exist?  A conservative  view would be that  it enables an
organism  to deviate  from  the  local gradient,  and  to better  than
locally optimize in the  struggle for self-presevervation, which would
require improvisation  under novel situations; once this  was done, FW
brought along with it potential by-products, like altruism, self-harm,
etc.,   behaviors    which   are   not    necessarily   conducive   to
self-preservation,  and  hence  not  `intended'  by  the  evolutionary
process that  gave rise to  FW. Another possibility  is that FW  is an
emergent phenomenon that is  a product of self-organization in complex
quantum physical systems.

Or, one can be more adventurous and regard FW as a primitive principle
in Nature,  apart from the  randomness and determinism.   We attribute
the  probabilistic  character  of  QM  to it,  since  a  deterministic
dynamics  could  not be  weakly  compatible  with  FW.  This  approach
provides John Wheeler's Really Big  Question "Why the quantum?" with a
teleological answer:  to accommodate FW.   Perhaps QM is  the simplest
indeterministic dynamics that can  accomodate FW.  To be precise, this
is applicable only to interpretations of QM that involve wave function
collapse. They are  most convenient as FW can  provide a mechanism for
collapse, and be weakly compatible with fundamental indeterminism that
collapse models  postulate.  Interesting models of  collapse have been
proposed by  various authors.  In  Ref. \cite{sri03}, we  suggest that
collapse  is not  a dynamical  but an  information  theoretic process,
precipitated by the build-up of sufficiently large entanglement, which
results in amplitudes becoming  unresolvable within the finite (though
in principle  unbounded) computational and  memory resources available
in Nature.

By  contrast,  in non-collapse  interpretations  of  QM,  such as  the
Many-Worlds Interpretation or  the Bohmian Interpretation \cite{bohm},
there  is  no  room  for  (non-illusory)  FW.  At  least  the  Bohmian
interpretation leaves open the possiblity that FW could interface with
the  deterministic equations  by influencing  the  initial probability
distribution, but there seems to  be no such redeeming feature for the
Many-worlds  interpretation \cite{gisin3440}.   It is  worth stressing
that the subjective  degrees of freedom that determine  the outcome of
wavefunction collapse should not  be confused with hidden variables in
the sense of  Bohm.  The latter represents a device to  turn QM into a
deterministic theory,  whereas in  our approach requires,  FW requires
indeterminism at the  objective level.  To the question  raised in the
title of this Section, then, we are able to respond: FW comes from the
same heaven that bestowed us with quantum randomness.

To  be aware  of a  theory  is to  be able  to  talk about  it and  to
understand its implications and limitations. To this end, one requires
a meta-theory, which is equipped with a langauge in which to formalize
propositions  about the  theory.   An axiomatic  system  that is  rich
enough to encompass the meta-theory will, in general, be more powerful
than  the one  that axiomatizes  the theory.  An instance  of  this is
provided   by   G\"odel's  theorem   \cite{god},   which  provides   a
metamathematical proof of mathematically undecidable propositions.

If  FW did  not exist,  and the  behavior of  observers  were entirely
determined  by the  rules  of  the base  theory,  then the  observers'
algorithmic complexity  \cite{chaitin} would not be  greater than that
of the theory, making  them incapable of encompassing the meta-theory.
Equipped with  FW, human experimenters  can be {\it  meta-entities} in
the theory. It allows them to freely pose questions, perform tests and
draw inferences about the theory  as external agents whose choices are
not entirely determined by the  theory. From this perspective, FW, may
be necessary  for observers to be  aware of a physical  theory such as
QM.   Future  work  should  make these  speculative  reflections  more
rigorous, and provide concrete  instances of cognitive phenomena being
seen in this light.  A somewhat related example is Ref.  \cite{sri08},
where we provide a concrete  instance where G\"odel's theorem may have
relevance for quantum gravity.

\section{Conclusions}

A  novel insight that  emerges here  is that  FW and,  more generally,
consciousness, may now  be amenable to quantitative analysis  in a way
that is in tune with the intuitive and philosophical approach to these
issues.  If we regard FW as a fundamental primitive and as an axiom in
a formalization of quantum  theory, then quantum indeterminism finds a
natural explanation.  It is worth  stressing that though FW provides a
mechanism for  wave function  collapse here, it  is not proposed  as a
means  to  resolve the  quantum  measurement  problem.  Our  preferred
approach to this problem was discussed in Section \ref{sec:godel}.

The possibility of FW-based  intervention, which depends on subjective
degrees of  freedom, implies that  QM is not universal.   The physical
actions  of  free-willed  agents  do  not have  an  entirely  physical
explanation: the world is not  closed under quantum physics.  Our work
shows  that  the  study  of   the  origin  and  existence  of  FW  has
ramifications for fields as diverse as neuroscience, computer science,
physics,   philosophy,   mathematical   logic  and   statistics.    It
potentially opens up many new directions of research.

The two  main immediate directions suggested  by our work  are, on the
experimental  side,  the neurobiological  tests  discussed in  Section
\ref{sec:ex},  and   on  the   theoretical  side,  development   of  a
mathematical model  for the mechanism  by which subjective  degrees of
freedom  couple to,  and are  able  to collapse,  the wavefunction  of
objective degrees  of freedom.  We  revisit this issue in  a following
work \cite{yyz}.


\begin{thebibliography}{99}
\bibitem{suControl} A. Suarez. Quantum randomness can be controlled by
  free   will–-  a  consequence   of  the   before-before  experiment.
  arxiv:0804.0871.
\end{thebibliography}
\end{document}